# On the Dynamics of a Small Rotating Particle Moving in Blackbody Radiation

A. A. Kyasov and G. V. Dedkov

Kabardino-Balkarian State University, Nanoscale Physics Group, Nalchik, Russia

Based on the general expressions for the tangential force and heating rate of a small polarizable particle during translational-rotational motion in the field of blackbody radiation (in the reference frame of latter), the tangential force in reference frame comoving with particle is calculated. It is this force determines the particle deceleration in the frame of reference of blackbody radiation.

## 1. Introduction

This short note is a continuation of our recent works devoted to the radiation and dynamics of a small polarizable particle in relative relativistic motion with respect to blackbody radiation. We consider a small spherical particle with radius $R$ and local temperature $T_1$ (in the own reference frame $\Sigma''$) moving with linear relativistic velocity $V$ relative to the radiation background of temperature $T_2$ (reference frame $\Sigma$) and rotating with angular velocity $\Omega \mathbf{n}$ in reference frame $\Sigma'$ comoving with particle, where $\mathbf{n}$ is the unit vector in the angular velocity direction and $\theta$ the angle between $\mathbf{n}$ and $V$ (Fig. 1). We also assume that the following conditions are satisfied: $\Omega R/c \ll 1$, $R \ll 2\pi \hbar c / k_B T_i$, $i = 1,2$, where $\hbar$, $k_B$, $c$ are the Planck's, Boltzmann's constants and the speed of light in vacuum. In this case, the particle can be treated as a point-like dipole (with fluctuating moments) when emitting low-frequency photons. Our modest aim is in obtaining an explicit formula for the frictional force acting on rotating particle in comoving reference frame $\Sigma'$. This allows one to calculate the particle deceleration in the reference frame $\Sigma$ of the blackbody radiation.

## 2. Results

In [1] (see also [3]), using general approach of fluctuation electrodynamics [4], we obtained the following dynamics equation of relativistic rotating particle moving in radiation background

$$\gamma^3 mc \frac{d\beta}{dt} = F_x - \frac{\beta}{1-\beta^2} \frac{\dot{Q}}{c} \tag{1}$$

$$F_x = -\frac{\hbar\gamma}{4\pi c^4}\int_{-\infty}^{+\infty}d\omega\,\omega^4\int_{-1}^{1}dx\,x \cdot$$

$$\left\{ \begin{array}{l} \left[(1-\beta^2)(1-x^2)\cos^2\theta + \left((1+\beta^2)(1+x^2)+4\beta x\right)\dfrac{\sin^2\theta}{2}\right] \cdot \\ \alpha''(\gamma\omega(1+\beta x))\left(\coth\dfrac{\hbar\omega}{2k_B T_2} - \coth\dfrac{\hbar\gamma\omega(1+\beta x)}{2k_B T_1}\right) + \\ + \left[(1-\beta^2)(1-x^2)\sin^2\theta + \left((1+\beta^2)(1+x^2)+4\beta x\right)\dfrac{1+\cos^2\theta}{2}\right] \cdot \\ \alpha''(\gamma\omega(1+\beta x)+\Omega)\left(\coth\dfrac{\hbar\omega}{2k_B T_2} - \coth\dfrac{\hbar(\gamma\omega(1+\beta x)+\Omega)}{2k_B T_1}\right) \end{array} \right\} \quad (2)$$

$$\dot{Q} = \frac{\hbar\gamma}{4\pi c^3}\int_{-\infty}^{+\infty}d\omega\,\omega^4\int_{-1}^{1}dx\,(1+\beta x) \cdot$$

$$\left\{ \begin{array}{l} \left[(1-\beta^2)(1-x^2)\cos^2\theta + \left((1+\beta^2)(1+x^2)+4\beta x\right)\dfrac{\sin^2\theta}{2}\right] \cdot \\ \alpha''(\gamma\omega(1+\beta x))\left(\coth\dfrac{\hbar\omega}{2k_B T_2} - \coth\dfrac{\hbar\gamma\omega(1+\beta x)}{2k_B T_1}\right) + \\ + \left[(1-\beta^2)(1-x^2)\sin^2\theta + \left((1+\beta^2)(1+x^2)+4\beta x\right)\dfrac{1+\cos^2\theta}{2}\right] \cdot \\ \alpha''(\gamma\omega(1+\beta x)+\Omega)\left(\coth\dfrac{\hbar\omega}{2k_B T_2} - \coth\dfrac{\hbar(\gamma\omega(1+\beta x)+\Omega)}{2k_B T_1}\right) \end{array} \right\} \quad (3)$$

where $\beta = V/c$, $\gamma = 1/\sqrt{1-\beta^2}$, and $\alpha''(\omega)$ is the imaginary part of the particle polarizability. It should be noted that tangential force $F_x$ and heating rate $\dot{Q}$ in Eqs. (1) – (3) correspond to the reference frame of blackbody radiation. The terms depending on $T_1$ and $T_2$ are related to the processes of emission and absorption of thermal photons. Equation (1) is valid irrespectively of the particle rotation, since the angular velocity enters in the right-hand side of (1) through $F_x$ and $\dot{Q}$ implicitly. In the case $\Omega = 0$, Eqs. (2) and (3) reduce to those obtained in [2].

The expression in the right-hand side of (1) is nothing more but the tangential force acting on the particle in co-moving reference frame $\Sigma'$, namely

$$F_x' = F_x - \frac{\beta}{1-\beta^2}\frac{\dot{Q}}{c}, \quad (4)$$

i. e. the dynamics equation takes the form

$$\frac{mc}{(1-\beta^2)^{3/2}}\frac{d\beta}{dt}=F'_x \quad (5)$$

Substituting (2), (3) into (4) and making use some transformations of the double integral yields

$$F'_x = \frac{\hbar}{4\pi c^4}\int_{-\infty}^{+\infty}d\omega\,\omega^4\int_{-1}^{1}dx\,x\coth\left(\frac{\hbar\omega\gamma(1+\beta x)}{2k_B T_2}\right)\cdot$$
$$\cdot\left[\begin{array}{l}\left[(1-x^2)\cos^2\theta+0.5(1+x^2)\sin^2\theta\right]\alpha''(\omega)+\\ +\left[(1-x^2)\sin^2\theta+0.5(1+x^2)(1+\cos^2\theta)\right]\alpha''(\omega+\Omega)\end{array}\right] \quad (6)$$

At $\Omega=0$, Eq. (6) reduces to

$$F'_x = \frac{\hbar}{4\pi c^4}\int_{-\infty}^{+\infty}d\omega\,\omega^4\int_{-1}^{1}dx\,x\,\alpha''(\omega)\coth\left(\frac{\hbar\omega\gamma(1+\beta x)}{2k_B T_2}\right) \quad (7)$$

From (7) it follows that $F'_x$ is always negative and corresponds to the frictional force, i. e. $d\beta/dt<0$ (from (5)). Equations (5) and (7) were first obtained in [2, 5]. The friction force in $\Sigma'$ does not depend on the particle temperature $T_1$ (both in (6) and (7)). This is a consequence of thermal radiation angular symmetry of particle in $\Sigma'$, which is not violated by rotation.

In the nonrelativistic limit $\beta\ll 1$, Eq. (6) reduces to

$$F'_x = -\frac{\hbar^2 V}{30\pi c^5 k_B T_2}\int_{-\infty}^{+\infty}\frac{d\omega\,\omega^5}{\sinh^2(\hbar\omega/2k_B T_2)}\left[(1+\sin^2\theta)\alpha''(\omega)+(3+\cos^2\theta)\alpha''(\omega+\Omega)\right] \quad (8)$$

Formula (8) generalizes the nonrelativistic formula by Mkrtchian et al. [6]:

$$F'_x = -\frac{\hbar^2 V}{3\pi c^5 k_B T_2}\int_0^{\infty}\frac{d\omega\,\omega^5\alpha''(\omega)}{\sinh^2(\hbar\omega/2k_B T_2)} \quad (9)$$

From (6) and (8) it follows that particle rotation may affect the dynamics of translational motion via the angle $\theta$ and the angular velocity $\Omega$. For example, if the particle has the resonance absorption at $\omega=\omega_0$, then

$$\alpha''(\omega) = \frac{\pi\alpha(0)\omega_0}{2}\left[\delta(\omega-\omega_0) - \delta(\omega+\omega_0)\right] \qquad (10)$$

Substituting (10) into (8) yields

$$F'_x = -\frac{\hbar V\alpha(0)\omega_0^5}{3c^5}\frac{x}{\sinh^2 x}\left[1 + \left(\frac{\Omega}{\omega_0}\right)^2(3+\cos^2\theta)\left(2 - 2x\coth x + 0.6x^2\coth^2 x - 0.2x^2\right)\right], \qquad (11)$$

where $x = \hbar\omega_0/2k_B T_2$. As we can see from (11), the impact of rotation is negligible in the case $\Omega \ll \omega_0$, but becomes significant with increasing $\Omega/\omega_0$. It is interesting to note that rotating particle may experience a small acceleration effect when the $\Omega$-dependent contribution to the friction force is positive. For example, at $\theta = 0$ and $x = 2.5$ the force $F'_x$ becomes accelerating (positive) at $\Omega/\omega_0 \geq 0.73$. Figure 2 shows the $F'_x$ (normalized to the factor $\frac{\hbar V\alpha(0)\omega_0^5}{3c^5}$) depending on the $\Omega/\omega_0$ and $x$ in the case $\theta = 0$, where the effect is maximal.

**Conclusion**

When a small polarizable rotating particle is moving in the blackbody radiation of certain temperature, its acceleration relative to the background is determined by the tangential force in the reference frame co-moving with particle. We have obtained general relativistic expression for this force depending on the particle polarizability, the temperature of background and vectors of linear and angular velocities. It is shown that in the nonrelativistic case under the condition of resonance absorption, it is possible that rotating particle experiences a small acceleration effect at some values of the background temperature $T_2$ and angular velocity $\Omega$.

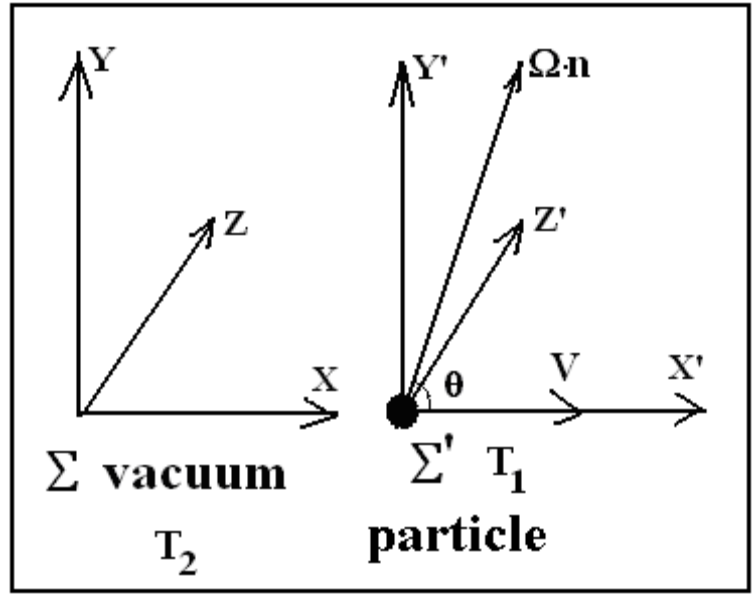

Fig. 1. Geometrical configuration and reference frames used. $\Sigma$: frame of reference of a blackbody radiation; $\Sigma'$: frame of reference comoving with particle; the own frame of reference of particle ($\Sigma''$) is not shown, its axis $Z''$ is directed along the angular velocity. Without loss of generality, the X' axis is chosen so that vector **n** lies in the X'Y' plane.

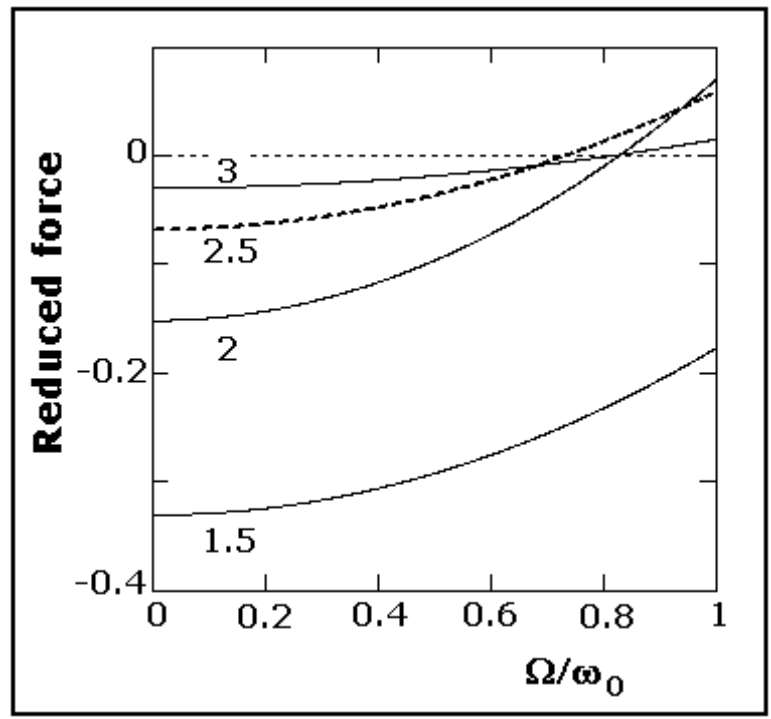

Fig.2. Reduced force $F'_x$, normalized to factor $\dfrac{\hbar V \alpha(0) \omega_0^5}{3c^5}$ (at $\theta = 0$). The lines from bottom to top are numbered according to the different values of $\hbar \omega_0 / 2k_B T_2$.